\documentclass[useAMS,usenatbib, amsmath, amssymb, a4paper,11pt]{article}

\usepackage{jcappub} 

\usepackage[T1]{fontenc} 
\usepackage{graphics}
\usepackage{color}

\def\vec#1{\mbox{\boldmath $#1$}}
\def\bmath#1{\mbox{\boldmath $#1$}}

\title{\boldmath Warping and tearing of misaligned circumbinary disks around eccentric SMBH binaries}

\author[a,1]{K.~Hayasaki,\note{Corresponding author.}}
\author[a,b]{B.~W.~Sohn,}
\author[c]{A.~T.~Okazaki,}
\author[a]{T.~Jung}
\author[a]{G.~Zhao}
\author[d]{and T.~Naito}

\affiliation[a]{Korea Astronomy and Space Science Institute, Daedeokdaero 776, Yuseong, Daejeon 305-348, Korea}
\affiliation[b]{Department of Astronomy and Space Science, University of Science and Technology, 217 Gajeong-ro, Daejeon, Korea}
\affiliation[c]{Faculty of Engineering, Hokkai-Gakuen University, Toyohira-ku, Sapporo 062-8605, Japan}
\affiliation[d]{Faculty of Management Information, Yamanashi Gakuin University, Kofu, Yamanashi 400-8575, Japan}

\emailAdd{kimi@kasi.re.kr}
\emailAdd{bwsohn@kasi.re.kr}
\emailAdd{okazaki@lst.hokkai-s-u.ac.jp}
\emailAdd{thjung@kasi.re.kr}
\emailAdd{gyzhao@kasi.re.kr}
\emailAdd{tsuguya@ygu.ac.jp}

\abstract{
We study the warping and tearing of a geometrically thin, non-self-gravitating disk surrounding binary 
supermassive black holes on an eccentric orbit. The circumbinary disk is significantly 
misaligned with the binary orbital plane, and is subject to the time-dependent tidal torques.
In principle, such a disk is warped and precesses, and is torn into mutually misaligned rings in the region, 
where the tidal precession torques are stronger than the local viscous torques. 
We derive the tidal-warp and tearing radii of the misaligned circumbinary disks 
around eccentric SMBH binaries. 
We find that in disks with the viscosity parameter $\alpha$ 
larger than a critical value depending on the disk aspect ratio, the disk warping 
appears outside the tearing radius. This condition 
is expressed as $\alpha > \sqrt{H/(3r)}$ for $H/r\lesssim0.1$, 
where $H$ is the disk scale height. 
If $\alpha < \sqrt{H/(3r)}$, only the disk tearing occurs because the tidal warp radius is 
inside the tearing radius, where most of disk material is likely to rapidly accrete onto 
SMBHs.
In warped and torn disks, both the tidal-warp and the tearing radii most strongly depend on the 
binary semi-major axis, although they also mildly depend on the other orbital and disk parameters. 
This strong dependence enables us to estimate the semi-major axis, once the tidal warp or tearing
radius is determined observationally: For the tidal warp radius of $0.1\,\rm{pc}$, the semi-major 
axis is estimated to be $\sim10^{-2}\,\rm{pc}$ for $10^7\,{\rm M}_\odot$ black hole with typical orbital 
and disk parameters. We also briefly discuss the possibility that central objects of observed warped 
maser disks in active galactic nuclei are supermassive black hole binaries.
}

\begin{document}
\maketitle
\flushbottom

\section{Introduction}
\label{sec:intro}
%
%
Supermassive black holes (SMBHs) with mass $10^5{\rm{M}}_\odot\lesssim{M}\lesssim10^{10}{\rm{M}}_\odot$ 
are considered to reside at the center of most galaxies \citep{kr95}. Hitherto, SMBHs have been 
found in 87 galaxies by observing the proper motion of stars bound by the SMBHs or by detecting 
radiation emitted from gas pulled gravitationally by the SMBHs \citep{kho13}. $\rm{H}_2\rm{O}$ 
maser emission from active galactic nuclei (AGNs) in spiral galaxies provides a strong tool to measure 
supermassive black hole (SMBH) masses, because it arises from a rotating disk on a subparsec scale 
with a nearly Keplerian velocity distribution around a SMBH. Those maser disks have been observed 
at the centers of NGC\,4258 \citep{miyoshi95}, NGC\,1068 \citep{green97}, NGC\,3079 \citep{yama04}, 
the Circinus galaxy \citep{green03}, UGC\,3789 \citep{reid09}, NGC\,6323 \citep{braatz07}, NGC\,2273, 
NGC\,6264, and some more objects \citep{kuo11}. Several maser disks show warped structure at the radii 
of $\sim0.1\,\rm{pc}$ \citep{green03,he05,kond08,kuo11}. From an observational point of view, maser spots 
on the disk in NGC~4258 are spatially distributed along a line on each side of a central black hole. The SMBH 
is then thought to be located at the center of a line connecting those two lines by a simple extrapolation, and 
the disk starts to warp at the innermost maser spot. What mechanism makes the disk warped still remains an 
open question.

%
%
Several scenarios have been proposed for explaining disk warping. \citet{pr96} showed that centrally illuminated 
accretion disks are unstable to warping due to the reaction force of reradiated radiation. Such a radiation-driven 
warping mechanism has also been applied to explain the disk warping in the context of X-ray binaries \citep{mb97,
wijers99,martin07,martin09}. If angular momentum vector of an accretion disk around a spinning black hole is 
misaligned with the spin axis, differential Lense-Thirring torque due to the frame-dragging effect aligns the inner 
part of the disk with the black-hole equatorial plane. Since the outer part of the disk retains its initial orientation, 
the resultant disk is warped \citep{bp75}. This Bardeen-Petterson effect is also considered to be a plausible 
mechanism for disk warping in maser disks \citep{caproni07}. Moreover, \citet{balex09} proposed that the warped 
disk at the center of NGC 4258 is caused by the process of resonant relaxation, which is a rapid relaxation mechanism 
to exchange angular momentum between the disk and the stars moving under the nearly spherical potential dominated 
by the SMBH. These mechanisms have been mostly discussed based on the assumption that the central object 
surrounded by the warped maser disk is a single SMBH.

%
%
The tight correlation between the mass of SMBHs and the mass or luminosity of the bulge of their host galaxies 
(\citep{mtr98,geb00,fm00,mm13}; see also \citep{kho13} for a review) suggests that the SMBH at the center of 
each galaxy should have evolved toward coalescence in a merged galaxy. If this is the case, a binary of SMBHs 
on a parsec or subparsec scale should be formed in a merged galactic nucleus before two black holes finally 
coalesce, yet no SMBH binaries have clearly been identified so far despite some claims (see \citep{ko06,po12} for 
reviews and references therein).

%
%
A possible link between the presence of SMBH binaries and the warping of observed maser disks
has been studied by \citet{khetal14} (hereafter, H14a).
They have, for simplicity, assumed that the circumbinary disk is initially aligned with the binary orbital plane. 
However, the angular momentum vector of the circumbinary disk does not always coincide with that of the binary orbital angular momentum, because the orientation of the circumbinary disk is primarily due to the angular momentum distribution of the gas supplied to the central region of AGNs. Therefore, the orientation of the circumbinary disk plane is expected to be distributed randomly with respect to the binary orbital plane. 
The formation of such a misaligned circumbinary disk around SMBH binaries has been numerically examined by \citet{dun14}. 
\citet{hsm13} have investigated the accretion process from misaligned circumbinary disks onto SMBHs in eccentric orbits.

%
%
Regarding the misaligned disk structure, the inner part of the circumbinary disk tends to align with the binary orbital plane, 
because the tidal precession torque is stronger than the shear viscous torque in the vertical 
direction, whereas the outer part tends to retain the original state because the tidal precession torque 
is weaker than the vertical viscous torque. This is the origin of disk warping \citep{nix11,flp13,lf13}.

%
%
On the other hand, \citet{nix13} have recently proposed that the circumbinary disk is broken into mutually misaligned rings, 
if the tidal precession torque is stronger than the local horizontal viscous torque in circular SMBH binaries. 
However, little is known about the relationship between disk tearing and warping in eccentric SMBH binaries.

%
%
In this paper, we examine the tidally driven warping and tearing of misaligned 
circumbinary disks around eccentric SMBH binaries. 
In Section~\ref{sec:2}, we describe the tidal torques originating from a time-dependent binary 
potential and derive the tidal warp radius of the misaligned circumbinary disk. In Section~\ref{sec:3}, 
we discuss a possibility that the observed warping of maser disks in several AGNs is caused by the 
tidal effect of SMBH binaries. Finally, Section~\ref{sec:4} summarises our scenario.

%
\section{Tidally driven warping and tearing of a misaligned circumbinary disk}
\label{sec:2}
%

\begin{figure}
\resizebox{\hsize}{!}{
\includegraphics{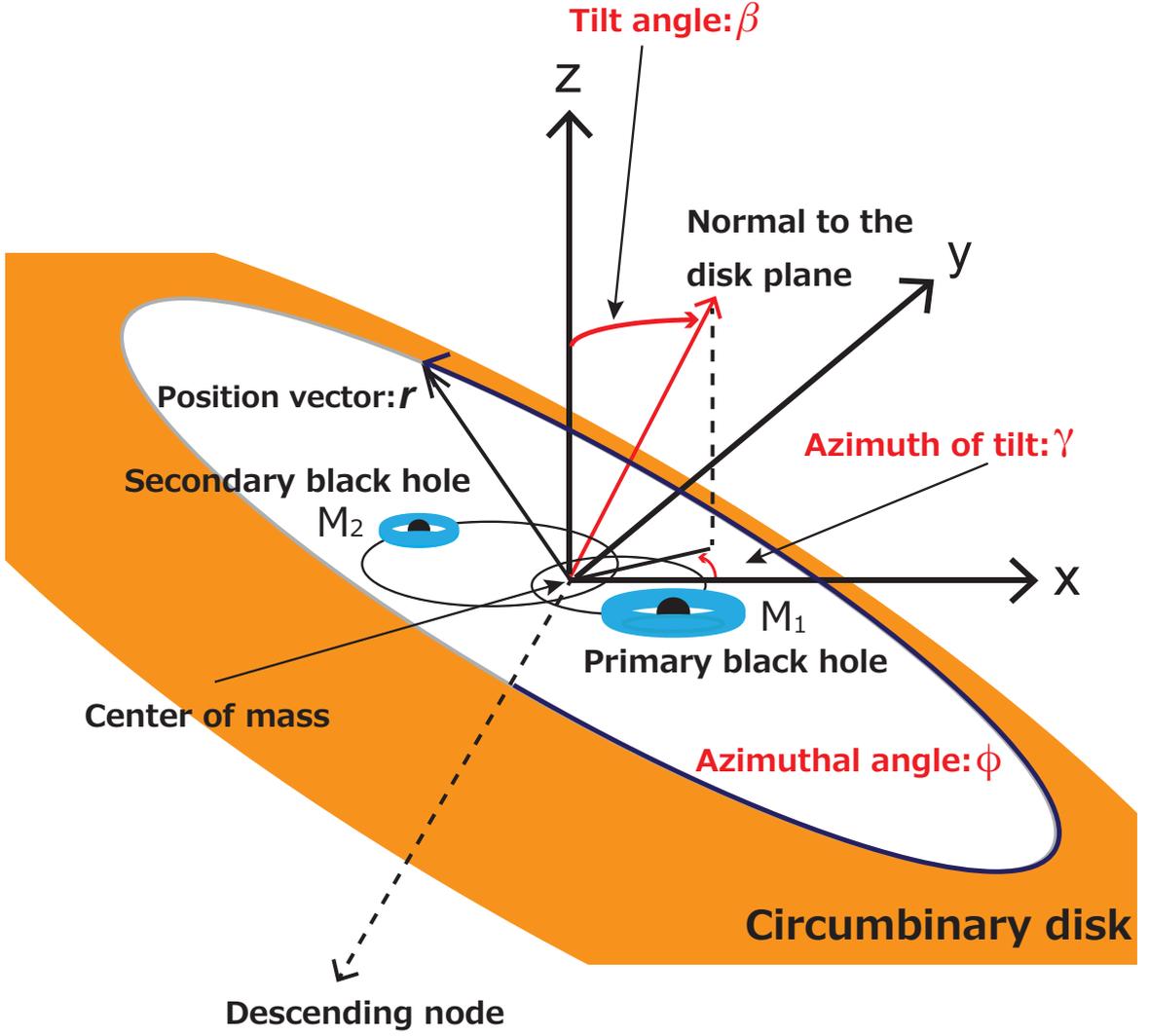}
}
\caption{
Configuration of a SMBH binary on an eccentric orbit and a circumbinary disk surrounding it. 
There are two angles ($\beta, \gamma$) which specify the orientation of the disk plane with 
respect to the binary orbital plane ($x$-$y$ plane). The azimuthal angle ($\phi$) of an arbitrary 
position on the disk is measured from the descending node.
}
\label{fig:schmatic}
\end{figure}

Let us consider the torques from the binary potential acting on the circumbinary disk, 
which is misaligned with the binary orbital plane, surrounding two black holes 
in a binary on a eccentric orbit. Figure~\ref{fig:schmatic} illustrates a schematic picture 
of the setting of our model; binary black holes orbiting each other are surrounded 
by a misaligned circumbinary disk.
The binary is put on the $x$-$y$ plane with its center of mass being at the origin 
in the Cartesian coordinate.
The masses of the primary and secondary black holes are represented 
by $M_1$ and $M_2$, respectively, and $M=M_1+M_2$. We put a circumbinary disk around the origin.
The unit vector of specific angular momentum of the disk is expressed 
by (e.g. \citet{pr96})
\begin{eqnarray}
\mbox{\boldmath $l$} 
=
\cos\gamma\sin\beta\hat{\boldmath{x}}
+ 
\sin\gamma\sin\beta
\hat{\boldmath{y}}
+
\cos\beta\hat{\boldmath{z}},
\label{eq:damvec}
\end{eqnarray}
where $\beta$ is the tilt angle between the circumbinary disk plane and the 
binary orbital plane, and $\gamma$ is the azimuth of tilt. 
Here, $\hat{\bmath{x}}$, $\hat{\bmath{y}}$, and $\hat{\bmath{z}}$ are unit 
vectors in the $x$, $y$, and $z$, respectively.
The position vector of the disk can be expressed by
\begin{eqnarray}
\mbox{\boldmath $r$}
&=&
r(\cos\phi\sin\gamma+\sin\phi\cos\gamma\cos\beta)\hat{\bmath{x}}
\nonumber \\
&+& 
r(\sin\phi\sin\gamma\cos\beta-\cos\phi\cos\gamma)\hat{\bmath{y}} 
\nonumber \\
&-&
r\sin\phi\sin\beta\hat{\bmath{z}}
\label{eq:rin}
\end{eqnarray}
where the azimuthal angle $\phi$ is measured from the descending node. 
The difference from equation~(3) of Paper~I is the position vector of each black hole, 
which is given by
\begin{equation}
\mbox{\boldmath $r$}_{i}=r_{i}\cos{f_i}\hat{\bmath{x}}+r_{i}\sin{f_i}\hat{\bmath{y}} \hspace{2mm}(i=1,2),
\label{eq:ri}
\end{equation}
where $f_2=f_1+\pi$ is the true anomaly and $r_{i}$ is written as
\begin{equation}
r_{i}=\xi_{i}\frac{a(1-e^2)}{1+e\cos{f_i}} 
\label{eq:rphi}
\end{equation}
with $\xi_1\equiv q/(1+q)$ and $\xi_2\equiv 1/(1+q)$. 
Here, $e$ is the binary orbital eccentricity, $q=M_2/M_1$ is the binary mass 
ratio, and $a=a_1 + a_2$ is the binary semi-major axis with $a_1\equiv\xi_1a$ 
and $a_2\equiv\xi_2a$. These and other model parameters are listed in Table~1.
We also assume that the disk total angular momentum is small compared with the 
binary angular momentum.

%
%
\begin{table*}
  \caption{Model parameters}
     \begin{tabular}{ll}
       \hline
       Definition & Symbol\\
       \hline
       Total black hole mass  & $M$  \\
       Primary black hole mass & $M_1$ \\
       Secondary black hole mass & $M_2$ \\
       Schwarzschild radius & $r_{\rm{S}}=2GM/c^2$ \\
       Binary mass ratio & $q=M_2/M_1$ \\
       Mass ratio parameter{s} & $\xi_1=q/(1+q)$, $\xi_2=1/(1+q)$ \\
       Binary semi-major axis & $a$ \\ 
       Orbital eccentricity & $e$ \\
       Orbital frequency & $\Omega_{\rm{orb}}=\sqrt{GM/a^3}$ \\
       Orbital period & $P_{\rm{orb}}=2\pi/\Omega_{\rm{orb}}$ \\
       True anomaly & $f_2=f_1+\pi$ \\  
       Tilt angle & $\beta$ \\
       Azimuth of tilt & $\gamma$ \\
       Azimuthal angle & $\phi$ \\
       Shakura-Sunyaev viscosity parameter & $\alpha$ \\
       Horizontal shear viscosity & $\nu_1$ \\
       Vertical shear viscosity & $\nu_2$\\
       Ratio of vertical to horizontal shear viscosities & $\eta=\nu_2/\nu_1$\\
       Mass-to-energy conversion efficiency & $\epsilon$ \\
       \hline
     \end{tabular}
   \label{tb:t1}
\end{table*}

%
%
%

%
\subsection{Tidal torques acting on the misaligned circumbinary disk}
%

The gravitational force on the unit mass at position $\bmath{r}$ 
on the circumbinary disk can be written by
\begin{eqnarray}
\bmath{F}_{\rm{grav}}=-\sum_{i=1}^{2}\frac{GM_i}{|\bmath{r}-\bmath{r}_i|^3}(\bmath{r}-\bmath{r}_i)
\end{eqnarray}
The corresponding torque is given by
\begin{eqnarray}
\bmath{t}_{\rm{grav}}=\bmath{r}\times\bmath{F}_{\rm{grav}}=\sum_{i=1}^2\frac{GM_i}{|\bmath{r}-\bmath{r}_i|^3}(\bmath{r}\times\bmath{r}_i)
\end{eqnarray}
We consider the tidal warping/precession with timescales much longer than 
local rotation period of the circumbinary disk. This allows us to use the torque 
averaged in the azimuthal direction and over the orbital period:
\begin{eqnarray}
&&
\langle\bmath{T_{\rm{grav}}}\rangle
\approx
\frac{1}{4\pi^2}\int_0^{2\pi}
\int_0^{2\pi}
\bmath{t}_{\rm{grav}}
\,d\phi
d(\Omega_{\rm{orb}}t)
\nonumber \\
&&
=
\frac{1}{2\pi}\int_0^{2\pi}
\left[\frac{1}{2\pi}\int_0^{2\pi}
\sum_{i=1}^2\frac{GM_i(\bmath{r}\times\bmath{r}_i)}{|\bmath{r}-\bmath{r}_i|^3}
d\phi\right]
\frac{(1-e^2)^{3/2}}{(1+e\cos{f}_i)^2}df
\nonumber \\
&=&
\frac{3}{8}\xi_1\xi_2
\frac{GM}{r}
\left(\frac{a}{r}\right)^2
\Biggr[
(1-e^2)
\sin\gamma\sin2\beta\hat{\bmath{x}}
-
(1+4e^2)
\cos\gamma\sin2\beta\hat{\bmath{y}}
\nonumber \\
&+&
5e^2\sin2\gamma\sin^2\beta\hat{\bmath{z}}
\Biggr],
\label{eq:tgrav}
\end{eqnarray}
which is equivalent to equation~(7) of \citet{khetal14b}, 
where $\Omega_{\rm{orb}}=\sqrt{GM/a^3}$ is the angular frequency of the mean binary motion.  
Here, we used for integration the following relationship:
\begin{eqnarray}
d(\Omega_{\rm{orb}}{t})=\frac{(1-e^2)^{3/2}}{(1+e\cos{f}_i)^{2}}df
\end{eqnarray}
and the approximations:
\begin{eqnarray}
|\bmath{r}-\bmath{r}_i|^{-3}&\approx&r^{-3}\left[1+3\frac{\bmath{r}\cdot\bmath{r}_i}{r^2}
+\mathcal{O}((r_i/r)^2)\right]
\label{eq:expan1}
\\
(1+e\cos{f}_i)^{-4}&\approx&1-4e\cos{f}_i+10e^2\cos^2{f}_i+\mathcal{O}(e^3),
\nonumber \\
\label{eq:expan2}
\end{eqnarray}
where $r\gg{a}$ is adopted for equations~(\ref{eq:expan1}).
The magnitude of the specific tidal torque is given by
\begin{eqnarray}
|\langle\bmath{T}_{\rm{grav}}\rangle|
&=&\frac{3}{8}\xi_1\xi_2
\frac{GM}{r}
\left(\frac{a}{r}\right)^2
\Theta(e,\beta,\gamma)
\label{eq:stgrav}
\end{eqnarray}
where $\Theta(e,\beta,\gamma)$ is a function given by
\begin{eqnarray}
\Theta(e,\beta,\gamma)
&
=
&
\Biggr
\{
4\biggr[
5e^2(2+3e^2)\cos^2\gamma
+
(e^2-1)^2
\biggr]
\cos^2\beta
\nonumber \\
&
+
&
25e^4\sin^22\gamma\sin^2\beta
\Biggr\}^{1/2}\sin\beta.
\label{eq:anglef}
\end{eqnarray} 
Figure~2 shows dependence of $\Theta(e,\beta,\gamma)$ on 
the three parameters: orbital eccentricity $e$, tilt angle $\beta$, 
and azimuth of tilt $\gamma$. 
It is noted from the figure that there are several zeros in 
$\Theta(e,\beta,\gamma)$, e.g.,  
for $\beta=0$ and $\pi$ with any $e$ and $\gamma$. 
Except for these values, $\Theta(e,\beta,\gamma)$ is of the order of one.
%
%
\begin{figure}
\resizebox{\hsize}{!}{
\includegraphics{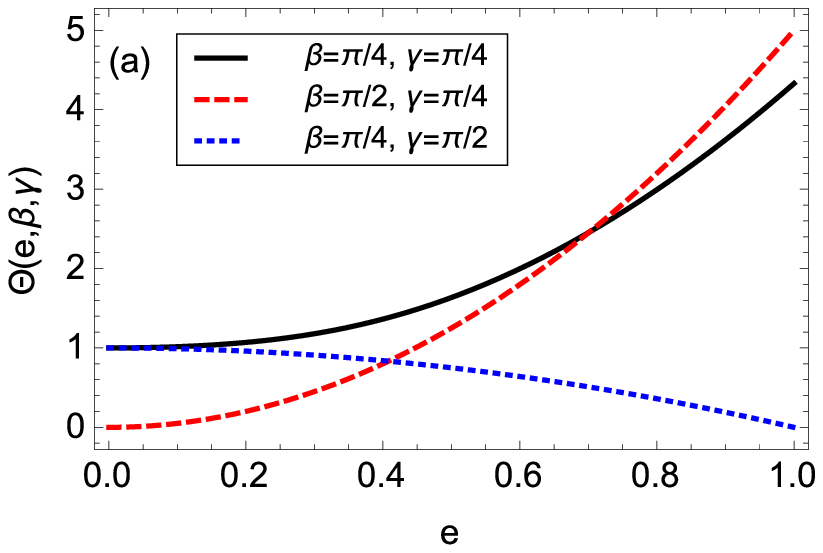}
\includegraphics{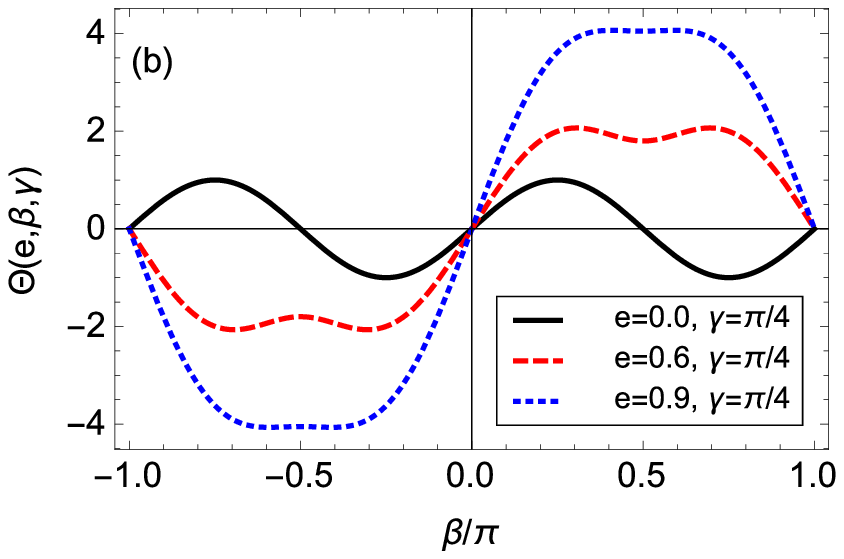}
\includegraphics{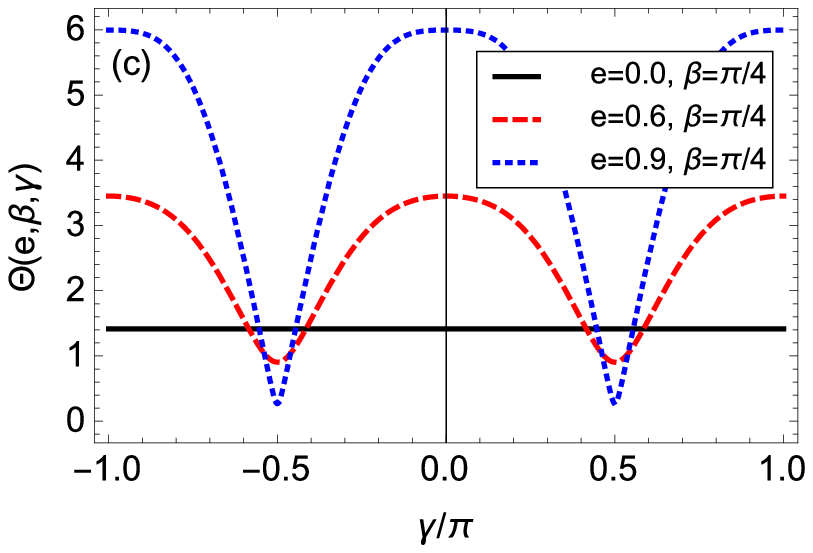}
}
\caption{
Dependence of $\Theta(e,\beta,\gamma)$ on orbital eccentricity 
$e$, tilt angle $\beta$, and azimuth of tilt $\gamma$.
Panel (a); dependence on $e$ with fixed values of $\beta$ and 
$\gamma$. The solid black, red dashed, and blue dotted lines indicate $\Theta(e,\pi/4,\pi/4)$, 
$\Theta(e,\pi/2,\pi/4)$, and $\Theta(e,\pi/4,\pi/2)$, respectively.
Panel (b); dependence on $\beta$ with fixed values of $e$ and $\gamma$.
The solid black, red dashed, and blue dotted lines indicate $\Theta(0.0,\beta,\pi/4)$, 
$\Theta(0.6,\beta,\pi/4)$, and $\Theta(0.9,\beta,\pi/4)$, respectively.
Panel (c); dependence on $\gamma$ with fixed values of $e$ and $\beta$.
The solid black, red dashed, and blue dotted lines indicate $\Theta(0.0,\pi/4,\gamma)$, 
$\Theta(0.6,\pi/4,\gamma)$, and $\Theta(0.9,\pi/4,\gamma)$, respectively.
}
\label{fig:timescale}
\end{figure}

%
%

By going through the same procedure as for equation~(\ref{eq:tgrav}), we obtain
the azimuthally-averaged and orbit-averaged tidal potential acting on the misaligned 
circumbinary disk around an eccentric binary:
\begin{eqnarray}
&&
\bar{\Phi}
=
\frac{1}{4\pi^2}\int_0^{2\pi}\int_0^{2\pi}
-\sum_{i=1}^2\frac{GM_i}{|\bmath{r}-\bmath{r}_i|}
d\phi{d(\Omega_{\rm{orb}}t)}
\nonumber \\
&\approx&
-\frac{GM}{r}\Biggr[
1
+
\frac{\xi_1\xi_2}{4}
\biggr[
1
+
\frac{3}{2}e^2
\biggr]
\left(\frac{a}{r}\right)^2
\nonumber \\
&-&
\frac{3\xi_1\xi_2}{8}
(
1-e^2+
5e^2\cos^2\gamma
)\sin^2\beta
\left(\frac{a}{r}\right)^2
\Biggr].
\label{eq:tpote}
\end{eqnarray}
Here, we used for integration equation~(\ref{eq:expan2})
and the following approximation:
\begin{eqnarray}
|\vec{r}-\vec{r}_i|^{-1}
\approx
r^{-1}\left[1+g_i\left(\frac{r_i}{r}\right)+\frac{1}{2}(3g_i^2-1)\left(\frac{r_i}{r}\right)^2+\mathcal{O}((r_i/r)^3)\right],
\label{eq:expan3}
\end{eqnarray}
where $g_i=\cos\phi\sin(\gamma-f_i)+\sin\phi\cos(\gamma-f_i)\cos\beta$.
Since the tidal potential has minima at $\beta=0$ (alignment) and $\beta=\pi$ (counter-alignment) 
for arbitrary $e$ and $\gamma$, 
the tidal torques tend to align or counter-align the tilted circumbinary disk with the orbital plane. 
Note that equation (\ref{eq:tpote}) is reduced to equation (3) of \citet{khato09} if $\beta=0$.

The tidal precession timescale for eccentric binaries is given by
\begin{eqnarray}
\tau_{\rm{prec}}&=&
\frac{
|\bmath{J}|
}
{
|\langle
\bmath{T}_{\rm{grav}}
\rangle|
}
=\frac{8}{3}\left(\frac{1}{\xi_1\xi_2}\right)\left(\frac{r}{a}\right)^{2}
\frac{1}{\Theta(e,\gamma,\beta)}
\frac{1}{\Omega}
\label{eq:tprec}
\\
&=&
\frac{4}{3\pi}\left(\frac{1}{\xi_1\xi_2}\right)\left(\frac{r}{a}\right)^{7/2}
\frac{1}{\Theta(e,\gamma,\beta)}
P_{\rm{orb}}
\nonumber 
\end{eqnarray}
where $\bmath{J}=r^2\Omega\bmath{l}$ with the disk angular frequency $\Omega=GM/r^3$ 
and $P_{\rm orb}=2\pi/\Omega_{\rm orb}$ are the specific angular momentum and binary orbital 
period, respectively.
Note that the tidal precession timescale depends on the tilt angle and azimuth of tilt for $e\neq0$. 
Since the inner edge of the disk is estimated to be $\sim2a$ \citep{al94}, 
the tidal precession timescale is longer than the binary orbital period.

%
%
\subsection{Tidal warping vs. tearing}
%
%

%
%
There are two types of important viscosities in the circumbinary disk.
The first type is the horizontal shear viscosity, $\nu_1$, which is the 
viscosity normally associated with accretion disks. The second type 
is the vertical shear viscosity, $\nu_2$, which tends to smooth out disk warping
when the disk is non-planar. \citet{og99} derived the relationship between $\nu_1$ and $\nu_2$:
\begin{eqnarray}
\eta=\frac{\nu_2}{\nu_1}=\frac{2(1+7\alpha^2)}{\alpha^2(4+\alpha^2)}
\label{eq:eta}
\end{eqnarray}
by taking a non-linear effect of the fluid on the warped disk,
where $\eta$ and $\alpha$ are the viscosity ratio parameter 
and the Shakura-Sunyaev viscosity parameter, respectively. 
For $\alpha\ll1$, the above equation is reduced to
\begin{eqnarray}
\eta\approx\frac{1}{2\alpha^2}.
\label{eq:eta2}
\end{eqnarray}

%
%
The global horizontal and vertical viscous timescales 
for a geometrically thin disk are estimated to be
\begin{eqnarray}
\tau_{\nu_1}&=&\frac{2}{3}\frac{r^2}{\nu_1}\approx\frac{2}{3}\frac{1}{\alpha}\frac{1}{\Omega}
\left(\frac{H}{r}\right)^{-2},
\\
\tau_{\nu_2}&=&\frac{2}{3}\frac{r^2}{\nu_2}\approx\frac{2}{3}\frac{1}{\eta\alpha}\frac{1}{\Omega}
\left(\frac{H}{r}\right)^{-2},
\end{eqnarray}
where $\nu_1\approx\alpha\Omega{H^2}$ with the disk scale hight $H$ and 
$\nu_2\approx\eta\alpha\Omega{H^2}$.
On the other hand, the local horizontal viscous timescale is estimated to be
\begin{eqnarray}
\Delta\tau_{\nu_1}&=&\frac{2}{3}\frac{r\Delta{r}}{\nu_1}\approx\frac{2}{3}\frac{1}{\alpha}\frac{1}{\Omega}
\left(\frac{H}{r}\right)^{-1},
\label{eq:localnu}
\end{eqnarray}
where $\Delta{r}\approx{H}$ is adopted.

%
%
The misaligned circumbinary disk is warped at a radius where the tidal precession timescale is longer than 
the vertical viscous timescale. We call this radius the tidal warp radius, following \citet{martin07,martin09}, 
who used it in the context of X-ray binaries. The tidal warp radius is given by 
\begin{eqnarray}
\frac{R_{\rm warp}}{r_{\rm{S}}}
&=&
\biggr[
\frac{3}{8}
\frac{\xi_1\xi_2}{\eta\alpha}
|\Theta(e,\beta,\gamma)|
\biggr]^{1/2}
\left(\frac{H}{r}\right)^{-1}
\left(\frac{a}{r_{\rm{S}}}\right),
\label{eq:rwarp}
\end{eqnarray}

\begin{figure}
\begin{center}
\resizebox{\hsize}{!}{\includegraphics{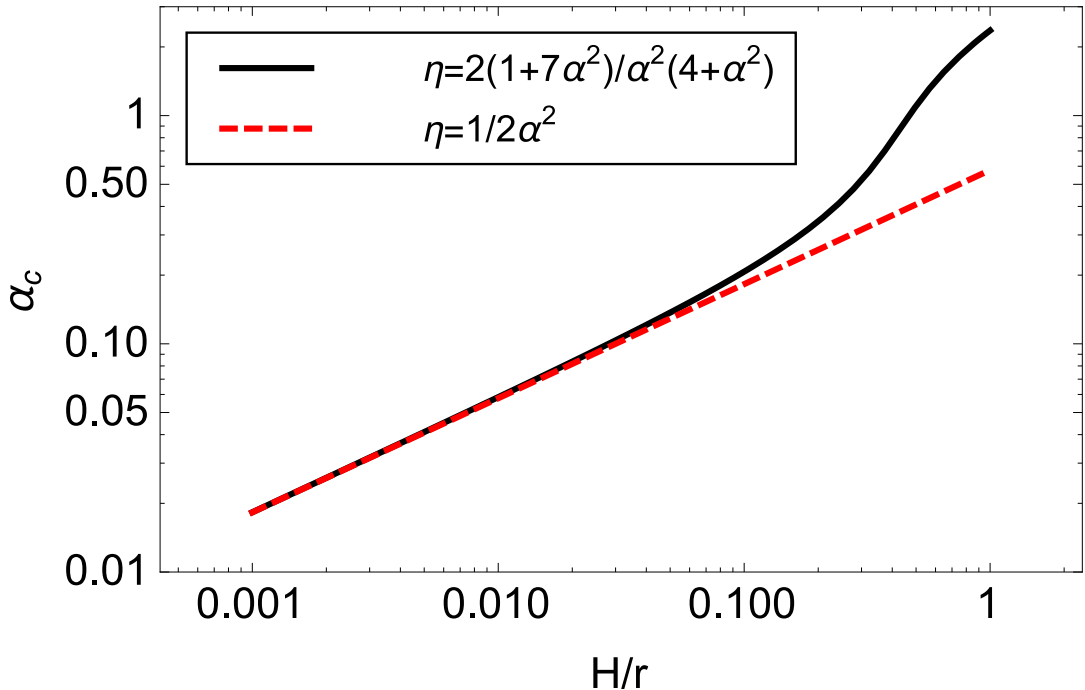}
}
\end{center}
\caption{
Critical value of viscosity parameter as a function of $H/r$.
The black solid line is the critical value, $\alpha_{\rm{c}}$, of $\alpha$ without approximation to $\eta$,  
whereas the red dashed line is $\alpha_{\rm{c}}$ for $\eta\approx1/(2\alpha^2)$.
}
\label{fig:alphac}
\end{figure}

Recently, \citet{nix13} have proposed that 
the circumbinary disk is torn into mutually 
misaligned gas rings if the 
timescale of the tidal precession, which is strongly differential, 
is shorter than the local horizontal viscous 
timescale. The disk tearing occurs at the radius: 
\begin{eqnarray}
\frac{R_{\rm tear}}{r_{\rm{S}}}
&=&
\biggr[
\frac{1}{4}
\frac{\xi_1\xi_2}{\alpha}
|\Theta(e,\beta,\gamma)|
\biggr]^{1/2}
\left(\frac{H}{r}\right)^{-1/2}
\left(\frac{a}{r_{\rm{S}}}\right).
\label{eq:rtear}
\end{eqnarray}
This reduces to equation (9) of \citet{nix13} when $e=0$.
The disk tearing is also confirmed by a three-dimensional 
Smoothed Particle Hydrodynamics simulations \citep{nix13}. 
Inside the tearing radius, the material rapidly accretes 
onto the central binary, especially, in a retrograde, misaligned 
circumbinary disk. Therefore, in order for the disk to have a warped 
structure, $R_{\rm{warp}}$ must be larger than $R_{\rm{tear}}$.
This condition is equivalent to
\begin{eqnarray}
\frac{3}{2}\frac{1}{\eta}
&>&
\frac{H}{r}.
\label{eq:crit1}
\end{eqnarray}
This is then rewritten as
\begin{eqnarray}
\alpha&>&\alpha_{\rm{c}},
\label{eq:crit2}
\end{eqnarray}
where $\alpha_c$ represents the critical value for the disk viscosity parameter, 
which is a function of $H/r$.
Figure~\ref{fig:alphac} shows the dependence of $\alpha_{\rm{c}}$ on $H/r$.
The red dashed line denotes the critical value of $\alpha$ for $\alpha\ll1$, 
which is equal to $\sqrt{H/(3r)}$ obtained by substituting equation~(\ref{eq:eta2}) 
into equation~(\ref{eq:crit1}). From the figure, we note $\alpha_{\rm{c}}=\sqrt{H/(3r)}$ is 
a good approximation for geometrically thin disks with $H/r<0.1$. For $\alpha>\alpha_{\rm{c}}$, the disk 
warping occurs outside the tearing radius. Otherwise, no disk warping will occur 
because there is little disk material at the tidal warp radius, which is inside the tearing radius.

\begin{figure}
\begin{center}
\resizebox{\hsize}{!}{\includegraphics{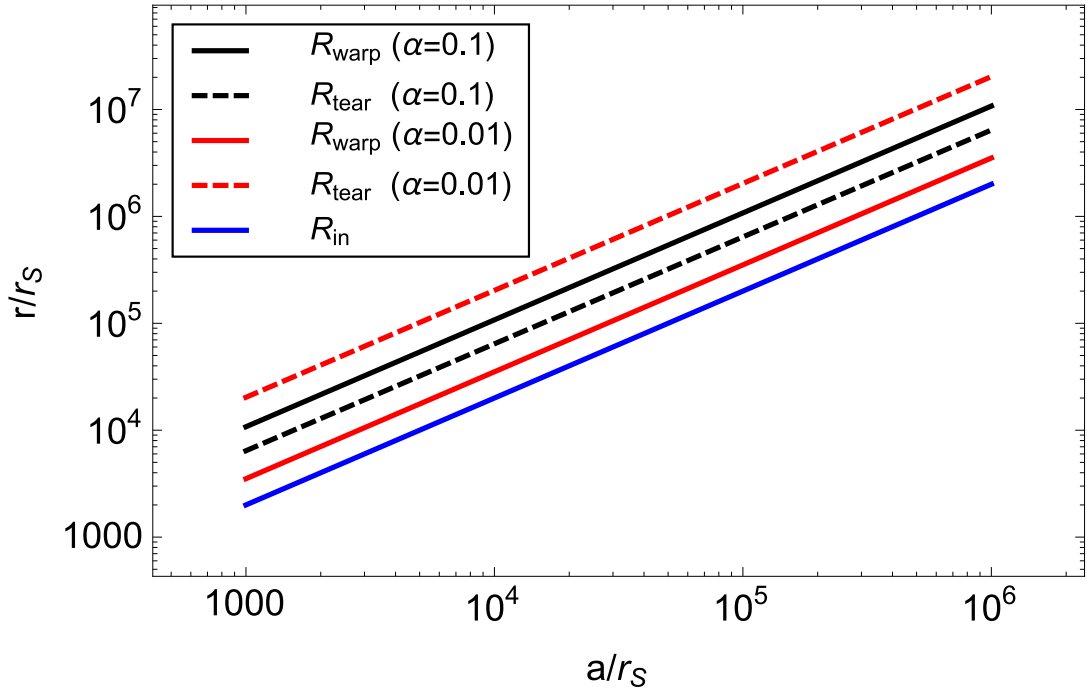}
}
\end{center}
\caption{
Dependence of tidal-warp and tearing radii on the semi-major axis.
The black solid and dashed lines are the tidal-warp and tearing radii 
for $\alpha>\alpha_{\rm{c}}$ case, respectively, whereas the red solid 
and dashed lines are the tidal-warp and tearing radii for $\alpha<\alpha_{\rm{c}}$ 
case, respectively. Here, $H/r=0.01\,(\alpha_{\rm{c}}\sim0.06)$ is adopted 
for each radius. The blue line shows the  inner edge radius of the circumbinary disk. 
The tidal-warp and tearing radii have other common parameters: 
$M=10^7M_\odot$, $q=0.1\,(\xi_1\xi_2=10/121)$, $e=0.6$, 
$\beta=\pi/4$, and $\gamma=\pi/4$, respectively.
}
\label{fig:wtradii}
\end{figure}

Figure~\ref{fig:wtradii} shows the dependence of tidal-warp and 
tearing radii on the semi-major axis. For a fixed value of $H/r=0.01$, 
$\alpha_{\rm{c}}\sim0.06$. We can confirm that the tidal warp radius 
is larger than the tearing radius for any semi-major axis for $\alpha=0.1$, 
whereas it is smaller than the tearing radius for any semi-major axis for $\alpha=0.01$.

%
%
%

%
%
%

The disk evolves diffusively for $\alpha > H/r$, whereas it does with dispersive 
wave propagation for $\alpha < H/r$ \citep{pp83}. For the latter case, 
the communication time due to wave propagation is roughly given by 
$2r/c_{\rm{s}}$, where $c_{\rm{s}}$ is the local sound speed. 
The tearing radius is then given by
\begin{eqnarray}
\frac{R_{\rm tear,w}}{r_{\rm{S}}}
&=&
\biggr[
\frac{3}{4}
\xi_1\xi_2
|\Theta(e,\beta,\gamma)|
\biggr]^{1/2}
\left(\frac{H}{r}\right)^{-1/2}
\left(\frac{a}{r_{\rm{S}}}\right).
\label{eq:rtear2}
\end{eqnarray} 
Comparing this equation with equation~(\ref{eq:rwarp}), 
we obtain, for $\alpha\ll1$,
\begin{eqnarray}
\frac{R_{\rm{warp}}}{R_{\rm{tear,w}}}
\approx
\left(\alpha\frac{r}{H}\right)^{1/2}
<1.
\end{eqnarray}
Thus, no disk warping occurs 
in a tilted circmbinary disk with $\alpha<{H/r}$.
%
%
\subsection{Warping and tearing of circumbinary disks in AGNs}
%
%

%
%

In this section, we apply our model to AGN disks.
A gaseous disk around a SMBH in an AGN is surrounded by a dusty torus (e.g. \citep{jaffe93}). 
The grains in the dusty torus is evaporated above the dust sublimation temperature by the irradiation emitted from the central source. The inner radius of the dusty torus should therefore be determined by the dust sublimation radius: $r_{\rm{dust}}=3\,{\rm{pc}}\,(L/10^{46}\,{\rm{erg\,s^{-1}}})^{1/2}(T_{\rm{dust}}/1500\,\rm{K})^{-2.8}$, where $T_{\rm{dust}}$ is the dust sublimation temperature \citep{bar87}. Assuming that the AGN luminosity is the Eddington luminosity: $L_{\rm{Edd}}\simeq1.3\times10^{38}(M/{\rm M}_\odot)\,\rm{erg\,s^{-1}}$, the dust sublimation radius is rewritten as $r_{\rm{dust}}=4.7\times10^{-1}\,(M/10^7\,{\rm{M}}_\odot)^{1/2}\,{\rm{pc}}$ with the adoption of $T_{\rm{dust}}=1500\,{\rm{K}}$. Since the circumbinary disk should be also inside the dusty torus in our scenario, 
we assume the disk outer radius and temperature at $R_{\rm{out}}$ to be given by
\begin{eqnarray}
\frac{R_{\rm{out}}}{r_{\rm{S}}}
\approx
4.8\times10^5
\left(
\frac{M}{10^7\,{\rm{M}}_\odot}
\right)^{-1/2}
\label{eq:rout}
\end{eqnarray}
and $T_{\rm{out}}=1500\,\rm{K}$, respectively.

%
%
Here, we assume that the radial profile of the disk temperature obeys 
\begin{eqnarray}
T=T_{\rm{out}}\left(\frac{r}{R_{\rm{out}}}\right)^s \hspace{4mm}\,(s<0),
\end{eqnarray}
where $s$, $R_{\rm{out}}$, and $T_{\rm{out}}$ are
the power law index of the radial profile of the disk temperature, disk outer radius, 
and temperature at $R_{\rm{out}}$, respectively. 
With the thin disk approximation, the ratio of the disk scale hight to radius is then given by
\begin{eqnarray}
\frac{H}{r}=
\sqrt{2}
\left(\frac{c_{\rm{s,out}}}{c}\right)
\left(\frac{R_{\rm{out}}}{r_{\rm{S}}}\right)^{1/2}
\left(\frac{r}{R_{\rm{out}}}\right)^{(s+1)/2},
\label{eq:hor}
\end{eqnarray}
where $c_{\rm{s,out}}=\sqrt{(R_{\rm{g}}/\mu)T}$ is the isothermal sound 
speed at $R_{\rm{out}}$ with the gas constant $R_{\rm{g}}$ and 
molecular weight $\mu$.

%
%
Substituting equation~(\ref{eq:hor}) into equation~(\ref{eq:rwarp}), 
the tidal warp radius is rewritten as
\begin{eqnarray}
\frac{R_{\rm warp,AGN}}{r_{\rm{S}}}
&=&
\biggr[
\frac{3}{16}
\frac{\xi_1\xi_2}{\eta\alpha}
\Theta(e,\beta,\gamma)
\biggr]^{1/(s+3)}
\left(\frac{c}{c_{\rm{s,out}}}\right)^{2/(s+3)}
\nonumber \\
&\times&
\left(\frac{R_{\rm{out}}}{r_{\rm{S}}}\right)^{s/(s+3)}\left(\frac{a}{r_{\rm{S}}}\right)^{2/(s+3)}.
\label{eq:rwagn}
\end{eqnarray}
Similarly, the tearing radius is rewritten as
\begin{eqnarray}
\frac{R_{\rm tear,AGN}}{r_{\rm{S}}}
&=&
\biggr[
\frac{1}{4\sqrt{2}}
\frac{\xi_1\xi_2}{\alpha}
\Theta(e,\beta,\gamma)
\biggr]^{2/(s+5)}
\left(\frac{c}{c_{\rm{s,out}}}\right)^{2/(s+5)}
\nonumber \\
&\times&
\left(\frac{R_{\rm{out}}}{r_{\rm{S}}}\right)^{s/(s+5)}
\left(\frac{a}{r_{\rm{S}}}\right)^{4/(s+5)}.
\label{eq:rtagn}
\end{eqnarray}

\begin{figure}
\begin{center}
\resizebox{\hsize}{!}{\includegraphics{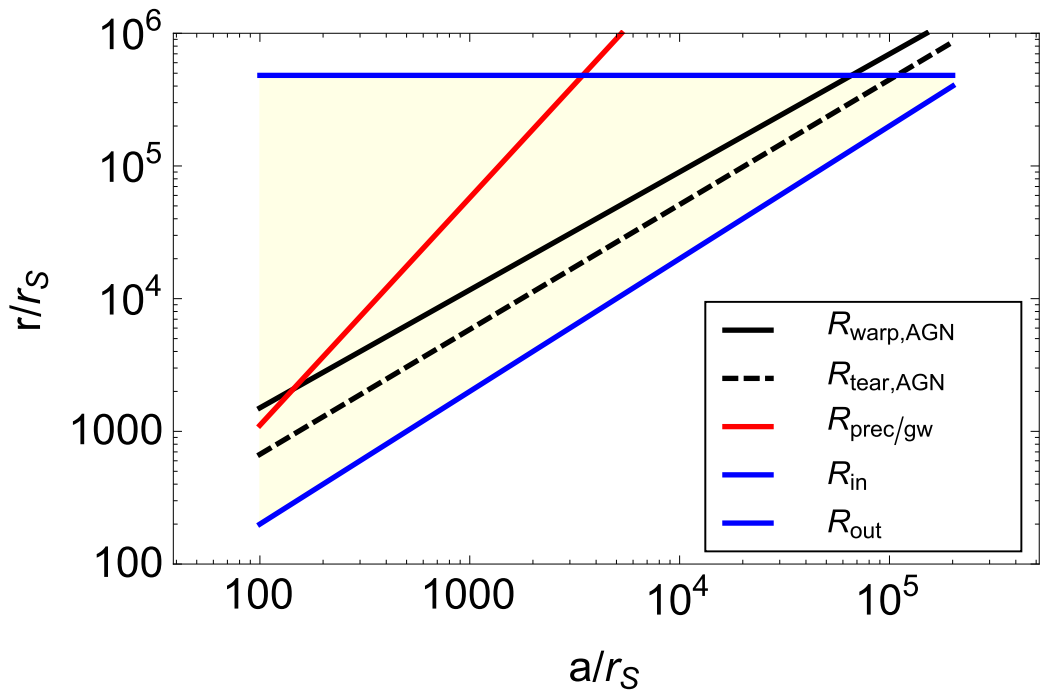}
}
\end{center}
\caption{
Characteristic radii of the warped and torn circumbinary disk around the SMBH binary 
on an eccentric orbit for $e=0.6$, $\alpha=0.1\,(\eta\approx50)$, $s=-3/4$, 
$q=0.1\,(\xi_1\xi_2=10/121)$, $M=10^7{\rm{M}}_\odot$, $\beta=\pi/4$, and 
$\gamma=\pi/4$. The solid and dashed black lines show the tidal warp and tearing 
radii, respectively. The red solid line shows the orbital decay radius where the tidal 
precession timescale equals the orbital decay timescale due to the GW emission 
(see equation~(\ref{eq:rprecgw})). While the blue lower line represents 
the inner radius of the circumbinary disk $r_{\rm{in}}\sim2a$, the blue upper line represents 
the outer radius of the circumbinary disk $R_{\rm{out}}/r_{\rm{S}}\approx4.8\times10^4\,(M/10^7\,{\rm{M}}_\odot)^{-1/2}$. 
The shaded area between the two blue lines represents the whole region of the circumbinary disk.
}
\label{fig:cradii}
\end{figure}

%
%
Figure~\ref{fig:cradii} shows characteristic radii of the misaligned circumbinary disk for $e=0.6$, 
$q=0.1\,(\xi_1\xi_2=10/121)$, $M=10^7{\rm{M}}_\odot$, $\alpha=0.1\,(\eta\approx50)$, 
$s=-3/4$, $\beta=\pi/4$, and $\gamma=\pi/4$. 
The solid and dashed black lines show the tidal warp and tearing radii, respectively. 
The red solid line shows the radius, which we call the orbital decay radius, where 
the tidal precession timescale equals the timescale in which the binary orbit decays by 
the gravitational wave (GW) emission. This orbital decay timescale for an 
eccentric binary is given by \citep{p64}
\begin{eqnarray}
\tau_{\rm{gw}}=\frac{5}{8}\frac{1}{\xi_1\xi_2}\left(\frac{a}{r_{\rm{S}}}\right)^{4}\frac{r_{\rm{S}}}{c}\frac{(1-e^2)^{7/2}}{1+73e^2/24+37e^4/96}
\label{eq:tgw}
\end{eqnarray}
Equating equation~(\ref{eq:tgw}) with equation~(\ref{eq:tprec}), we obtain the orbital decay radius:
\begin{eqnarray}
\frac{R_{\rm{prec/gw}}}{r_{\rm{S}}}
&=&
\left[
\frac{15}{64\sqrt{2}}
\frac{\Theta(e,\beta,\gamma)(1-e^2)^{7/2}}{1+73e^2/24+37e^4/96}
\right]^{2/7}
\left(\frac{a}{r_{\rm{S}}}\right)^{12/7}.
\label{eq:rprecgw}
\end{eqnarray}
When it is larger than the tidal warp and tearing radii, the circumbinary 
disk can be warped and torn before two SMBHs coalesce. Otherwise, two SMBHs coalesce prior 
to disk warping and tearing. 
The blue solid lower and upper lines show the inner and outer radii of the circumbinary disk, respectively. 
The shaded area between the two blue lines shows the whole region of the circumbinary disk.

%
%
%

%
\section{Discussion}
\label{sec:3}
%

\begin{figure}
\begin{center}
\resizebox{\hsize}{!}{\includegraphics{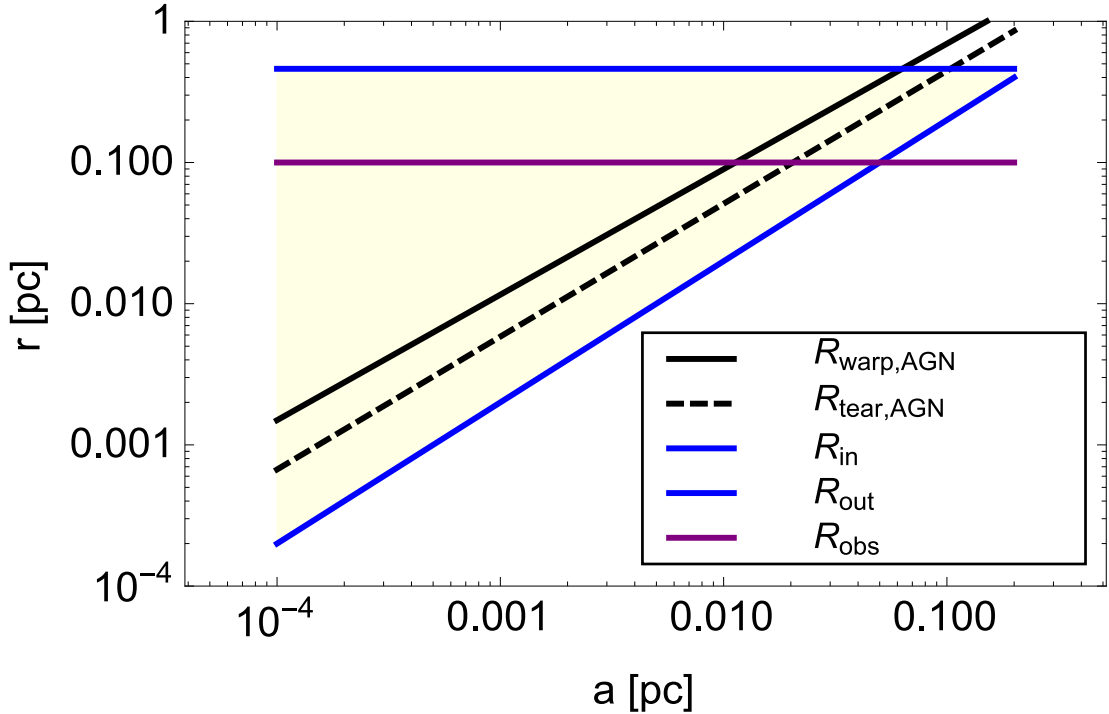}
}
\end{center}
\caption{
Estimate of a semi-major axis of a SMBH binary 
by the relationship between the tidal warp and tearing 
radii and the observed radius, which is shown by the 
purple solid line. The solid and dashed black lines are, 
respectively, the tidal-warp and tearing radii of the misaligned 
circumbinary disk around the eccentric SMBH binary with 
$e=0.6$, $\alpha=0.1\,(\eta\approx50)$, $s=-3/4$, $q=0.1\,(\xi_1\xi_2=10/121)$, 
$M=10^7{\rm{M}}_\odot$, $\beta=\pi/4$, and $\gamma=\pi/4$.
The shaded area between the two blue lines represents the 
whole region of the circumbinary disk.
}
\label{fig:semi}
\end{figure}

%
%
In this section, we discuss a possible link between the observational signatures 
of circumbinary disk warping and tearing and the presence of SMBH binaries.
If the warp and/or tearing radii are observationally determined,
we can estimate the semi-major axis of the SMBH binary from equations~(\ref{eq:rwagn}) and (\ref{eq:rtagn}) as
\begin{eqnarray}
\frac{a_{\rm{warp}}}{r_{\rm{S}}}
&=&
\Biggr[
\frac{16}{3}
\left(\frac{\eta\alpha}{\xi_1\xi_2}\right)
\frac{1}{\Theta(e,\beta,\gamma)}
\Biggr]^{1/2}
\left(\frac{c_{\rm{s,out}}}{c}\right)
\left(\frac{R_{\rm{out}}}{r_{\rm{S}}}\right)^{-s/2}
\nonumber \\
&\times&
\left(\frac{R_{\rm{obs}}}{r_{\rm{S}}}\right)^{(s+3)/2},
\label{eq:semiw}
\\
\frac{a_{\rm{tear}}}{r_{\rm{S}}}
&=&
\Biggr[
4\sqrt{2}
\left(\frac{\alpha}{\xi_1\xi_2}\right)
\frac{1}{\Theta(e,\beta,\gamma)}
\Biggr]^{1/2}
\left(\frac{c_{\rm{s,out}}}{c}\right)^{1/2}
\left(\frac{R_{\rm{out}}}{r_{\rm{S}}}\right)^{-s/4}
\nonumber \\
&\times&
\left(\frac{R_{\rm{obs}}}{r_{\rm{S}}}\right)^{(s+3)/4},
\label{eq:semitear}
\end{eqnarray}
where $R_{\rm{obs}}$ shows the tidal warp or tearing radius determined by observations.
Figure~\ref{fig:semi} shows the relationship between this observed radius 
of disk warping or tearing and the binary semi-major axis. 
Here, we adopt $\alpha=0.1\,(\eta\approx50)$, $s=-3/4$, $q=0.1\,(\xi_1\xi_2=10/121)$, 
$\beta=\pi/4$, and $\gamma=\pi/4$.
The estimated semi-major axes are $a_{\rm{warp}}\sim0.01\,\rm{pc}$ 
and $a_{\rm{tear}}\sim0.02\,\rm{pc}$, although they mildly depend on the orbital eccentricity, 
binary mass ratio, tilt angle, and azimuth of tilt, except for $\beta=\pi/2, 0$. 
Because it is observationally more difficult to distinguish between 
the disk warping and tearing, the actual
semi-major axis is likely to be $a_{\rm{warp}}\lesssim{a}\lesssim{a_{\rm{tear}}}$.

%
%
There are observational evidences for disk warping in the maser disks at the center of NGC 4258 
\citep{he05}, Circinus \citep{green03}, NGC\,2273, UGC\,3789, NGC\,6264, and NGC\,6323 \citep{kuo11}. 
The tidally-driven warping or tearing of misaligned circumbinary disks can be one of mechanisms 
for explaining the warped structure of the maser disks in these systems. Since no observational 
information is currently available about the warping or tearing radius, we simply assume that these 
maser disks start to be warped or torn at the innermost maser spot radii. Here, we pick up NGC\,4258 
case because of the most remarkable example of disk warping among the known maser disk systems. 
The inner most maser spot radius is $0.17\,\rm{pc}$. Assuming that $R_{\rm{obs}}=0.17\,\rm{pc}$, 
the semi-major axis is estimated to be $1.2\times10^{-2}\,\rm{pc}$.
However, it is difficult to distinguish, solely by the current analysis, whether the central object 
is a single SMBH or such a small-scale SMBH binary. Other independent theoretical and observational approaches are needed.

%
%
Probing GWs from individual SMBH binaries with masses $\gtrsim10^7{\rm{M}}_\odot$ 
with pulser timing arrays (PTAs; \citep{lb01,svv09}) gives a powerful tool to 
determine if the central object surrounded by the warped maser disk is a SMBH binary
or a single SMBH. For a typical PTA error box ($\approx40\,\rm{deg}^2$) in the sky, the 
number of interloping AGNs are of the order of $10^2$ for more than $10^8\,{\rm M}_\odot$ 
black holes if the redshift range is between $0$ and $0.8$ (see Figure 1 of \citet{tk12} in detail).
As discussed in H14a, even if the observed warped maser disk systems host SMBH binaries, 
the characteristic amplitudes of GWs from these SMBH binaries are three to four orders of magnitude 
smaller than the current PTA sensitivity, so that it is unlikely for GWs to be detected from these systems.

%
%
%

%
\section{Conclusion}
\label{sec:4}
%

%
%
We have investigated the tidally driven warping and tearing of a geometrically thin, non-self-gravitating 
circumbinary disk around two SMBHs in a binary on an eccentric orbit, where the original disk plane is misaligned with 
the binary orbital plane. While tidal torques acting on the tilted circumbinary 
disk tend to align it with the orbital plane, the viscous torque in the vertical direction tends 
to retain its original orientation. The disk can then be warped at the tidal warp radius, 
where the tidal precession timescale equals the vertical viscous timescale. 
On the other hand, the disk is broken into mutually misaligned rings inside the tearing 
radius, where the tidal precession timescale equals the local horizontal 
viscous timescale.

There is a critical value of disk viscosity parameter.
If $\alpha>\sqrt{H/(3r)}$ with $H/r\lesssim0.1$, 
the tidal warp radius is larger than the tearing radius. 
Then, the disk tearing first occurs and subsequently disk warping 
occurs outside the tearing radius. If $\alpha<\sqrt{H/(3r)}$ with $H/r\lesssim0.1$, only 
the disk tearing occurs, because the disk material inside the 
tearing radius is likely to accrete rapidly.

The tidal-warp and tearing radii most strongly depend on the binary 
semi-major axis, although it mildly depends on the other orbital and disk parameters 
except for $\beta=0\,,\pi$. This strong dependence enables us to estimate the 
semi-major axis, once the tidal-warp or tearing radius is determined observationally. 
For the tidal-warp or tearing radius of $0.1\,\rm{pc}$, for instance, 
SMBH binaries with masses of $10^{7}{\rm{M}}_\odot$ and other typical 
orbital and disk parameters are estimated to have a binary separation on a 
$10^{-2}\,\rm{pc}$ scale.


\acknowledgments
They also thank Christopher~Nixon for helpful suggestions.
K.H. is grateful to Jongsoo~Kim for helpful discussions and his continuous encouragement. 
K.H. would also like to thank the Kavli Institute for Theoretical Physics (KITP) for their hospitality 
and support during the program on A Universe of Black Holes. B.W.S. and T.H.J. are grateful for 
support from KASI-Yonsei DRC program of Korea Research Council of Fundamental Science 
and Technology (DRC-12-2-KASI). This work was also supported in part by the Grants-in-Aid for 
Scientific Research (C) of Japan Society for the Promotion of Science (23540271 T.N. and K.H.; 
24540235 A.T.O. and K.H.).








%

\end{document}